\documentclass{ws-procs975x65}

\begin{document}

\title{On the Frame Fixing in Quantum Gravity}

\author{S. Mercuri}

\address{ICRA --- International Center for Relativistic Astrophysics}

\author{G. Montani}

\address{ICRA --- International Center for Relativistic Astrophysics}
\address{Dipartimento di Fisica, Universit\`a di Roma ``La Sapienza'', 
Piazzale Aldo Moro 5, I-00185, Roma, Italy}

\maketitle

\abstracts{We provide a discussion about the necessity to fix the reference
frame before quantizing the gravitational field. Our presentation is based on
stressing how the 3+1-slicing of the space time becomes an ambiguous procedure
as referred to a quantum 4-metric.}

In the Wheeler-DeWitt (WDW) approach \cite{DeW1967}, the quantization of gravity is performed in the canonical way, starting from the Arnowitt-Deser-Misner (ADM) action. The use of the ADM
formalism\cite{ArnDesMis1959} 
is justified by the necessity to obtain Hamiltonian constraints, but the
straightforward quantization of such (3+1)-picture contains some relevant
ambiguities. In fact, the aim of the WDW approach is to quantize the
gravitational field in a particular representation and its outcoming provides
essentially information on the quantum dynamics of the 3-metric tensor
defined on spatial hypersurfaces. \newline 
To use the ADM splitting is equivalent to a kind of ``gauge fixing'', because it is preserved only under restricted coordinates transformations (time displacements and 3-diffeomorphisms); the point here is that the ``gauge fixing'' depends on the field we are quantizing and therefore the canonical approach seems to be an ambiguous procedure.


Since in the ADM action the conjugate momenta, $\pi$ and $\pi^{i}$, respectively to the lapse function $N$ and to the shift vector $N^{i}$ are
constrained to vanish, then, on a quantum level, the wave
functional of the system does not depend on the lapse function and on the shift
vector. The ambiguity relies on regarding as equivalent the fully covariant approach  and the ``gauge fixed'' ADM one, in fact passing from
$g_{\mu\nu}$ to ADM variables involves a metric dependent procedure, in
the sense that we must be able to define a unit time-like normal field
$n^{\mu}$ $(g_{\mu\nu}n^{\mu}n^{\nu}=-1),$ which ensures the space-like nature
of $h_{ij}$ (in this respect we recall that $h_{ij}\equiv g_{\mu\nu}%
\partial_{i}y^{\mu}\partial_{j}y^{\nu}$ corresponds to the spatial components
of the 4-tensor $h_{\mu\nu}=g_{\mu\nu}+n_{\mu}n_{\nu}).$ Now the following
question arises: \emph{how is it possible to speak of a unit time-like normal
field for a quantum space-time?} Indeed such a notion can be recognized, in
quantum regime, at most in the sense of expectation values; therefore assuming the existence of $n^{\mu}$ before quantizing the system dynamics makes
the WDW approach physically ill defined.

Our point of view is that the canonical quantization of the gravitational
field can be performed in a (3+1)-picture only if we add, to such a
scheme, some information about the existence of the time-like normal field, as
shown in \cite{Mon2002,MerMon2003a}, this result can be achieved by including in the dynamics
the \emph{kinematical action} \cite{Kuc1981}, already adopted to quantize
``matter'' fields on a fixed background \cite{Kuc1981}. The physical interpretation of such
new term either on a classical as well as on a quantum level leads to
recognize the existence of a reference fluid and in this sense the analysis of \cite{Mon2002,MerMon2003a,MerMon2003b} converges with the literature on the frame fixing problem (see \cite{BicKuc1997} and references therein). We observe that to include the kinematical
action can be regarded as a consequence of fixing in the gravity action the lapse function and the shift vector and, therefore, to choose four independent components of the
gravitational field, which is just the outcoming of the frame fixing. 

A more physical manner to ensure the existence of a time-like vector consists
of filling the space time with a fluid which plays the role of real reference
frame. Here we discuss on a phenomenological ground, the canonical
quantization of the gravitational field plus a dust reference fluid, outlining
some relevant differences between the classical and quantum behavior of this
system. \newline The Einstein equations and the conservation law, for the
coupled gravity-fluid system, take the form
\begin{equation}
G_{\mu\nu}=\chi\varepsilon u_{\mu}u_{\nu},\label{xy1}
\quad 
u^{\nu}\nabla_{\nu}u^{\mu}=0,
\quad \nabla_{\nu}(\varepsilon u^{\nu})=0,
\end{equation}
where $G_{\mu\nu}$ and $\chi$ denotes respectively the Einstein tensor and
constant.\newline Remembering a well-known result, it is easy to show that the
following relations take place \cite{Thi2001}
\begin{equation}
G_{\mu\nu}u^{\mu}u^{\nu}=-\frac{H(h_{ij},p^{ij})}{2\sqrt{h}}%
=\chi\varepsilon,\label{xy3}
\quad G_{\mu\nu}u^{\mu}h_{i}^{\nu}=\frac{H_{i}(h_{ij},p^{ij})}{2\sqrt{h}}=0.
\end{equation}
Here $h_{ij}$ ($ij=1,2,3$) denotes the 3-metric of the spatial hypersurfaces
orthogonal to $u^{\mu}$ and $p^{ij}$ its conjugate momenta, while $H$ and
$H_{i}$ refer respectively to the super-Hamiltonian and to the super-momentum of
the gravitational field. The above relations hold if we make reasonable assumption that the conjugate momentum $p^{ij}$ is not affected by the matter variables (i.e. the fluid term
in ADM formalism should not contain the time derivative of the 3-metric
tensor). Only the Hamiltonian constraints are
relevant for the quantization procedure and, in the comoving frame, when the
4-velocity becomes $u^{\mu}=\{1,\mathbf{{0}\}}$ ($N=1\;N^{i}=0$), we have to
retain also the conservation law
$\varepsilon\sqrt{h}=-{\omega(x^{i})}/{2\chi},\label{xy5}$
where $h\equiv deth_{ij}$ and $x^{i}$ denote the spatial coordinates of the
comoving frame. Indeed, a crucial point in the above considerations relies on
the synchronous nature of the comoving frame as consequence of the geodesic
motion of the dust fluid. \newline Thus, when the coordinates system becomes a
real physical frame, the Hamiltonian constraints read
\begin{equation}
H=\omega(x^{i})\;\quad\quad\quad H_{i}=0\,.\label{xy6}%
\end{equation}
Now, to assign a Cauchy problem for such a system, for which equations
(\ref{xy6}) play the role of constraints on the Cauchy data, corresponds to
provide on a (non-singular) space-like hypersurface, say $\Sigma^{(0)}$, the
values $\{h_{ij}^{(0)},p^{(0)ij},\varepsilon^{(0)}\}$; from these values $\omega^{(0)}$
can be calculated by (\ref{xy6}).\newline It follows that, by
specifying a suitable initial condition, the value of $\omega^{(0)}$ can be
made arbitrarily small; from the constraints point of view, a very small value
of $\omega^{(0)}$ means, if $h^{(0)}$ is not so, that the fluid becomes a test
one (being $\omega$ a constant of the motion); we emphasize that for finite
values of $\omega$, $h$ should not vanish to avoid unphysical diverging energy
density of the fluid.

The canonical quantization of this system is achieved as soon as we implement
the canonical variables into quantum operators and annihilate the state
functional $\Psi$ via the Hamiltonian operator constraints. Thus the quantum
dynamics obeys the following eigenvalue problem:
\begin{equation}
\widehat{H}\Psi(\{h_{ij}\},\omega)=\omega\Psi(\{h_{ij}\},\omega),\label{xy7}%
\end{equation}
where $\{h_{ij}\}$ refers to a whole class of 3-geometries, so that
the super-momentum constraint holds automatically.\newline We stress how the above result is equivalent to the eigenvalues problem
obtained in \cite{Mon2002}. In the above equation (\ref{xy7}), the spatial function $\omega$ plays the
role of the super-Hamiltonian eigenvalue; in this respect, we observe how
its values can no longer be assigned by the initial values, but they
have to be determined via the spectrum of $\widehat{H}$. We conclude that, in the
quantum regime, a real dust reference fluid never approaches a test system.\newline Moreover the presence of non zero eigenvalues for the super-Hamiltonian removes the so called ``frozen formalism'' of the WDW equation and confirms the idea that introducing a physical unit time like vector provides a consistent and evolutive canonical quantum gravity dynamics.

\end{document}